\documentclass[12pt,preprint]{aastex}
\usepackage{graphicx,lscape}

\newcommand{\xmm}{{\em XMM-Newton}}

\newcommand{\swift}{{\em Swift}}

\newcommand{\src}{J0556}

\newcommand{\nh}{N\ensuremath{_{\rm H} }}	
\newcommand{\msun}{M\ensuremath{_\odot}}	
\newcommand{\about}{\ensuremath{\sim}}		

\newcommand{\nvi}{N\,{\sc vi}}
\newcommand{\nvii}{N\,{\sc vii}}
\newcommand{\oviii}{O\,{\sc viii}}
\newcommand{\NOsolarll}{27} 
\newcommand{\NOsolarval}{57} 
\newcommand{\NOll}{3.6} 
\newcommand{\NOval}{7.5} 

\shorttitle{Strong 24.8 \AA\ emission line in \src}
\shortauthors{Maitra et al.}

\begin{document}
\title{
A strong emission line near 24.8 angstrom in the X-ray binary
system MAXI~J0556--332: gravitational redshift or unusual donor?
}

\author{
  Dipankar Maitra\altaffilmark{1}, 
  Jon M. Miller\altaffilmark{1},
  John C. Raymond\altaffilmark{2}, and
  Mark T. Reynolds\altaffilmark{1}
}
\affil{Department of Astronomy, University of Michigan,
    Ann Arbor, MI 48109, USA}
\affil{Harvard-Smithsonian Center for Astrophysics, 
    60 Garden Street, Cambridge, MA 02138, USA}
\email{dmaitra@umich.edu}

\begin{abstract}   
We report the discovery of a strong emission line near 24.8 \AA\
(0.5 keV) in the newly discovered X-ray binary system MAXI~J0556-332
with the reflection grating spectrometer onboard the \xmm\ observatory.
The X-ray light curve morphology during these observations is complex
and shows occasional dipping behavior. Here we present time- and
rate-selected spectra from the RGS and show that this strong emission
line is unambiguously present in all the XMM observations.  The measured
line center is consistent with the Ly-$\alpha$ transition of \nvii\ in
the rest frame. While the spectra contain imprints of absorption lines
and edges, there appear to be no other significantly prominent narrow
line due to the source itself, thus making the identification of the 24.8
\AA\ line uncertain. We discuss possible physical scenarios, including
a gravitationally redshifted \oviii\ Ly-$\alpha$ line originating at
the surface of a neutron star or an unusual donor with an extremely high
N/O abundance ($>$\NOsolarval) relative to solar, that may have produced
this comparatively strong emission line.
\end{abstract}


\section{Introduction} \label{s:intro} 
Monitor of All-sky X-ray Image (MAXI) detected a new X-ray source on
2011 January 11, subsequently named MAXI~J0556-332 \citep[][hereafter
we will refer to the source as \src]{atel3102}.  The discovery was
immediately followed up by pointed observations in other wavelengths.
X-ray observations made with \swift\ indicated that the interstellar
extinction toward \src\ is small (\nh$\sim$10$^{21}$ cm$^{-2}$;
\citealt{atel3103}) compared to most Galactic X-ray binaries (XRBs),
which typically have \nh\about10$^{22}$ cm$^{-2}$.  The low \nh, largely
owing to the high Galactic latitude of the source ($b$=-25.183), implies
that minimal line-of-sight absorption will facilitate efforts to study
its low-energy X-ray spectrum.  The X-ray spectral and timing properties
of \src\ suggest that it is an XRB currently going through an episode
of active accretion.  The optical counterpart for this source was found
to be a stellar object with USNO B1.0 magnitudes of $m_R$=19.91 and
$m_B$=19.52, and an optical spectrum obtained by \citet{atel3104} showed
emission lines from H$\alpha$, He I, and He II, indicating the presence
of an accretion disk near the compact accretor.  Spaceborne X-ray and
groundbased I-band monitoring (using the CTIO/SMARTS 1.3m telescope)
shows that the source is still active as of 2011 October 20.

The nature of the compact accretor or the companion donor star in
\src\ is currently unknown, although most evidence point towards a
neutron star (NS) accretor.  No ``type 1'' bursts have been seen from
the source so far, but the X-ray timing properties and evolution on
color-color diagram are similar to those of NS XRBs \citep{atel3650}.
The optical-to-X-ray and radio-to-X-ray flux ratios are also similar
to Galactic NS XRBs \citep{atel3116,atel3119}.  The distance to the
source is also currently unknown.  Analyzing the archival data of the
source from the Catalina Real-Time Transient Survey revealed that no
optical outburst had been detected since August 2005 \citep{atel3328}.
Data from {\it RXTE/All Sky Monitor} show that this is its first X-ray
outburst since January 1996.

\section{Data Reduction and Analysis} \label{s:data} 
\xmm\ observed \src\ on  MJD 55608 and 55653 (hereafter referred
to as Obs1 and Obs2 respectively), each observation being of
about 40 ksec duration.  We analyzed the data obtained by the
reflection grating spectrometer \citep[RGS;][]{denHerder+2001}
using \xmm\ Science Analysis Software (SAS; v.11.0.0) and
following the standard extraction procedures outlined in the \xmm\ User
Guide\footnote{http://xmm.esac.esa.int/external/xmm\_user\_support/documentation/sas\_usg/USG/}.
The background light curves were examined for the presence of proton
flares.  While there were no flares during Obs1, data from the initial
14 ksec of Obs2 had to be discarded due to high background.  The SAS
extraction regions for source and background spectra were inspected
using the spatial- and energy-dispersion plots and left unchanged from
the SAS default.  The observed fluxes in each RGS CCD were compared
with their 2\% pile-up flux limits (\S3.4.4.8.1 of the \xmm\ User's
Handbook), and we found that pile-up was significantly less than 2\%
even for the steady-high state (when the source was brightest among the
two observations).

\begin{figure} 
\centering
\includegraphics[angle=-90, width=1.0\textwidth]{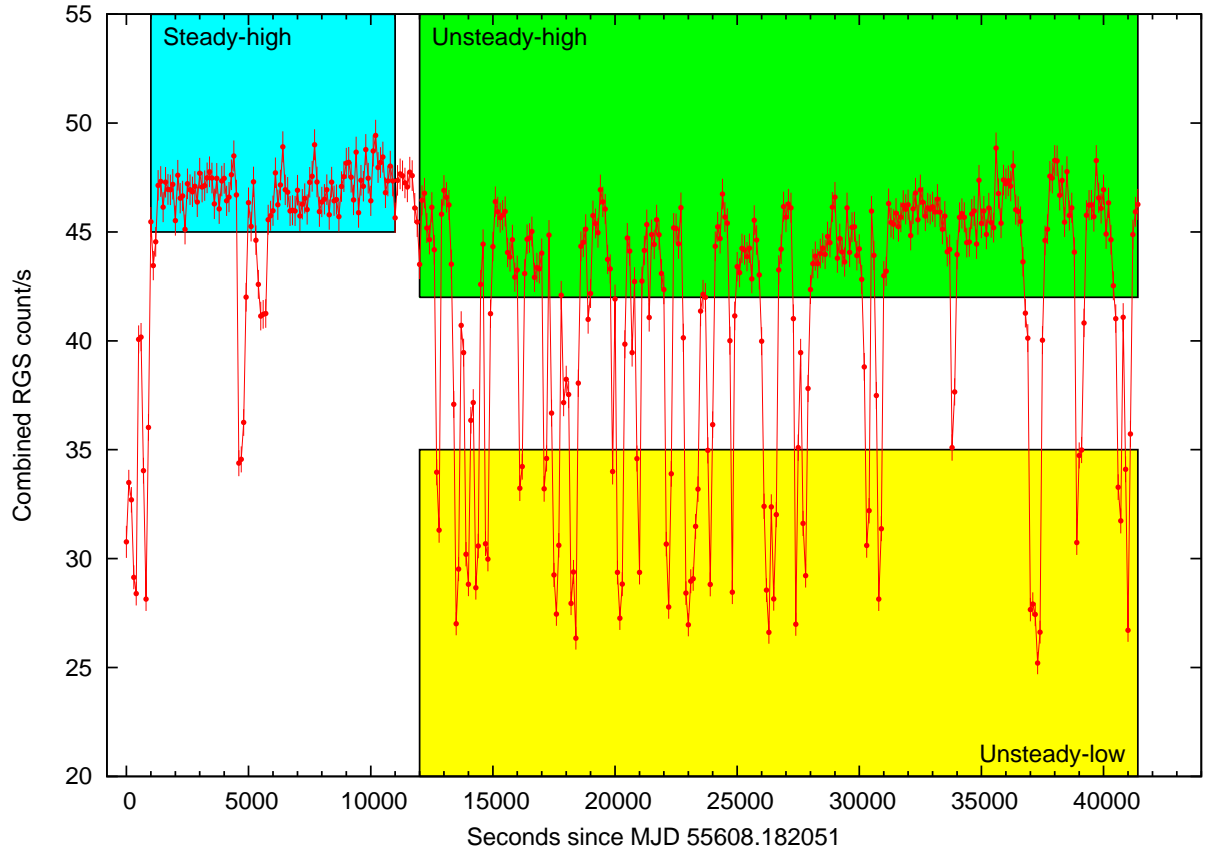}
\caption[]{%
RGS light curve of the first \xmm\ observation (Obs1) showing its complex 
morphology, especially epochs of dipping and non-dipping behavior.  The 
variously colored regions represent the intervals from which
spectra were analysed.
}
\label{f:lc}
\end{figure}

The average RGS count-rates during Obs1 and Obs2 were 41.2 counts/s
and 12.1 counts/s respectively.  The light curve morphology during Obs1
(Fig.~\ref{f:lc}) is quite complex, showing dipping behavior for \about20
ksec during the middle of the observation as well as extended epochs
without any dips. This confirms previous shorter X-ray observations
of \src\ where the dips were seen occasionally \citep[e.g.  as reported
by][]{atel3110,atel3112}. Given the marked changes in the source behavior,
it is likely that the spectral properties were changing during Obs1.
Therefore we extracted spectra from 3 separate intervals based on the
combined RGS count-rate ($cps$) and time ($t$ seconds since the start
of the observation): (1) {\it steady-high} -- when 1,000$<$t$<$11,000
and $cps$$>$45, which constitutes most of the time when the source flux
was steady; (2) {\it unsteady-high} -- when t$>$12,000 and $cps$$>$42,
selecting the photons collected during the peaks of the dipping
period, and (3) {\it unsteady-low} -- when t$>$12,000 and $cps$$>$35,
selecting the photons collected during the minima of the dipping period.
These selections are marked in Fig.~\ref{f:lc} by differently colored
regions. Once the background flare was excluded, the Obs2 light curve was
steady and did not show any dips or any other signs of large variability.
Therefore we created one single time-averaged spectrum of the entire
Obs2 after excluding the initial flare.

We analyzed data from the full RGS bandpass (7--36\AA) for both RGS1
and RGS2 units.  Since the second order RGS spectra do not cover the
wavelength region near the 24.8 \AA\ line, we restrict our analysis to
the first order spectra only.  The spectra were grouped for all fits
with the criterion that there were at least 30 counts/bin.  Fits to the
X-ray spectra were performed in wavelength space using ISIS (v.1.6.1-36;
\citealt{isisref}) and XSPEC (v12.7.0; \citealt{xspecref}).  The solar
abundance table given by \citet{angr} were used for all fits.

\begin{figure} 
\centering
\includegraphics[angle=0, width=1.0\textwidth]{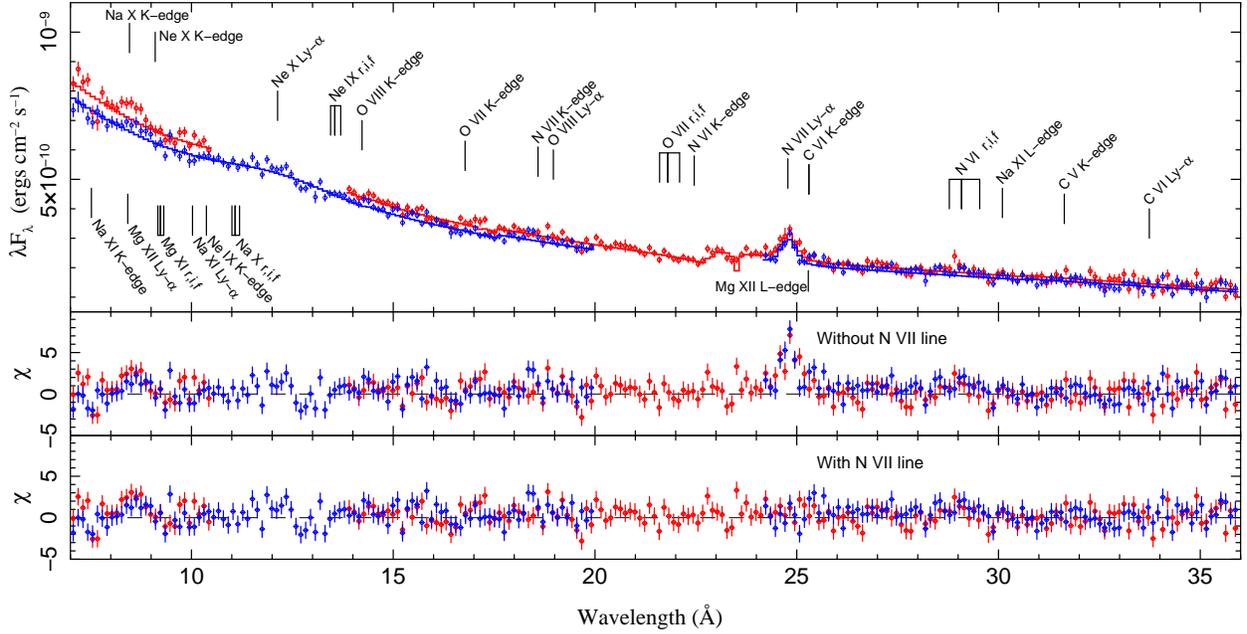}
\caption[]{%
\xmm/RGS spectrum of \src, in 7--36 \AA\ range, obtained on MJD
55608 during the {\it steady-high} period (i.e. the cyan region in
Fig.~\ref{f:lc}).
{\em Top panel}: RGS1 data in red and the RGS2 data in blue (binned
for visual clarity only). Data from both RGS units were fitted jointly
as described in text. The joint best-fit models for RGS1 and RGS2 are
shown by red and blue histograms respectively.  Rest-frame energies of
few different lines/edges typically prominent in this range are labeled.
The energies of the resonance (r), intercombination (i), and forbidden
(f) lines are also shown.  The broad feature in the 10--14 \AA\ range
was modeled with a Gaussian centered at 12.2 \AA\ and width ($\sigma$)
of 1.4 \AA.
{\em Middle panel}: Residuals in units of standard deviation, when
the normalization of the narrow line at 24.8 \AA\ (0.5 keV) was 
set to zero.
{\em Bottom panel}: Residuals in units of standard deviation, to
the best-fit continuum + line model.
}
\label{f:line}
\end{figure}

Apart from the 24.8 \AA\ emission line which is readily visible in the RGS
spectra (Fig.~\ref{f:line}), there are also many low-significance ``bumps
and wiggles''.  However there appear to be no strong narrow spectral
features associated to the source that are of comparable strength to the
24.8 \AA\ line.  Signatures of the intervening interstellar medium (ISM)
are seen via the presence of neutral and ionized atomic oxygen edges and
absorption lines in \about22.1--23.8 \AA\ region.  Also evident in the
brighter observations is the instrumental neutral O K-edge absorption
feature near 23.5 \AA\ as noted, e.g., by \citet{denHerder+2001}.
There is a hint of an excess near 29 \AA, which may be due to a
blend of He-like \nvi\ resonance line (28.78 \AA\ in rest frame) and
intercombination lines (29.082 and 29.084 \AA).  Note that the possible
\nvi\ feature does not show the forbidden component at 29.535 \AA. But
that is not surprising because the emission region in a compact XRB would
be expected to be much denser than 10$^{10}$ cm$^{-3}$.  Also, if the
excess is indeed due to \nvi, this would support a nitrogen-rich donor
scenario.  While there are no other narrow lines of comparable strength,
there is a broad feature near 12.2 \AA\ ($\sim$1 keV; covered only by
RGS2 and not by RGS1).  The feature can be well modeled by a Gaussian
centered near 12.2 \AA\ and a width of 1.4 \AA. Such a broad feature near
$\sim$1 keV is sometimes seen in XRBs \citep[see, e.g.,][]{Cackett+2010}.
Given the quality of the data, we cannot rule out the possibility that
it could be instrumental, or relativistically broadened Ne X line,
or smeared Fe L-edge seen in reflection.

Various low-significance features are not only difficult to identify
but also make the determination of the continuum difficult.  For all
the observations, the RGS continuum can be well modeled with a
phenomenological broken power law model.  We also attempted to model
the continuum with a thermal (both blackbody as well as accretion disk
models were tried) plus power law model.  However the thermal component
in these models was not constrained and the best-fit temperatures
were much higher than that of the energy range considered.  Thus in
effect the thermal+power law models were mimicking a broken power law.
The exact choice of continuum in the comparatively narrower RGS energy
range is not crucial for measuring the 24.8 \AA\ line parameters, and
we stress that the choice of the broken power law model is nothing
but an empirical description of the continuum for the purpose of
measuring the line.  A full analysis of the 0.6--10 keV EPIC data is
beyond the scope and focus of the current {\it Letter}, and will be
presented in a later work (Maitra et al., in prep.).  However we would
briefly like to note that the phenomenological thermal+power law fits
to the PN data shows no strong evidence for hardening of the power
law component during the dips.  The power law photon index during
dips (unsteady-low) is $\Gamma$=2.65$\pm$0.10 and that outside dips
(unsteady-high) is 2.51$\pm$0.06.  However, both the temperature and
normalization of the thermal component are smaller during the dips (when
kT$_{BB}$=0.70$\pm$0.15 keV, N$_{BB}$=90$\pm$21) when compared to those
outside the dips (kT$_{BB}$= 0.81$\pm$0.02 keV, N$_{BB}$= 143$\pm$13).
As a result, the ratio of unabsorbed non-thermal-to-thermal flux
within the PN energy range is higher during the dips compared to that
outside the dips.  We modeled absorption of X-rays by the ISM using
the Tuebingen-Boulder model (\texttt{tbnew}; an updated version of
\citealt{Wilms+2000}).  The best-fit model parameters for the various
observations/selections are given in Tables~\ref{t:cont_fits} and
\ref{t:line_fits}.  The quoted errors for the best-fit models correspond
to a 90\% confidence limit for the continuum model parameters, and 68\%
confidence limit for line parameters.

\begin{figure} 
\centering
\includegraphics[angle=-90, width=1.0\textwidth]{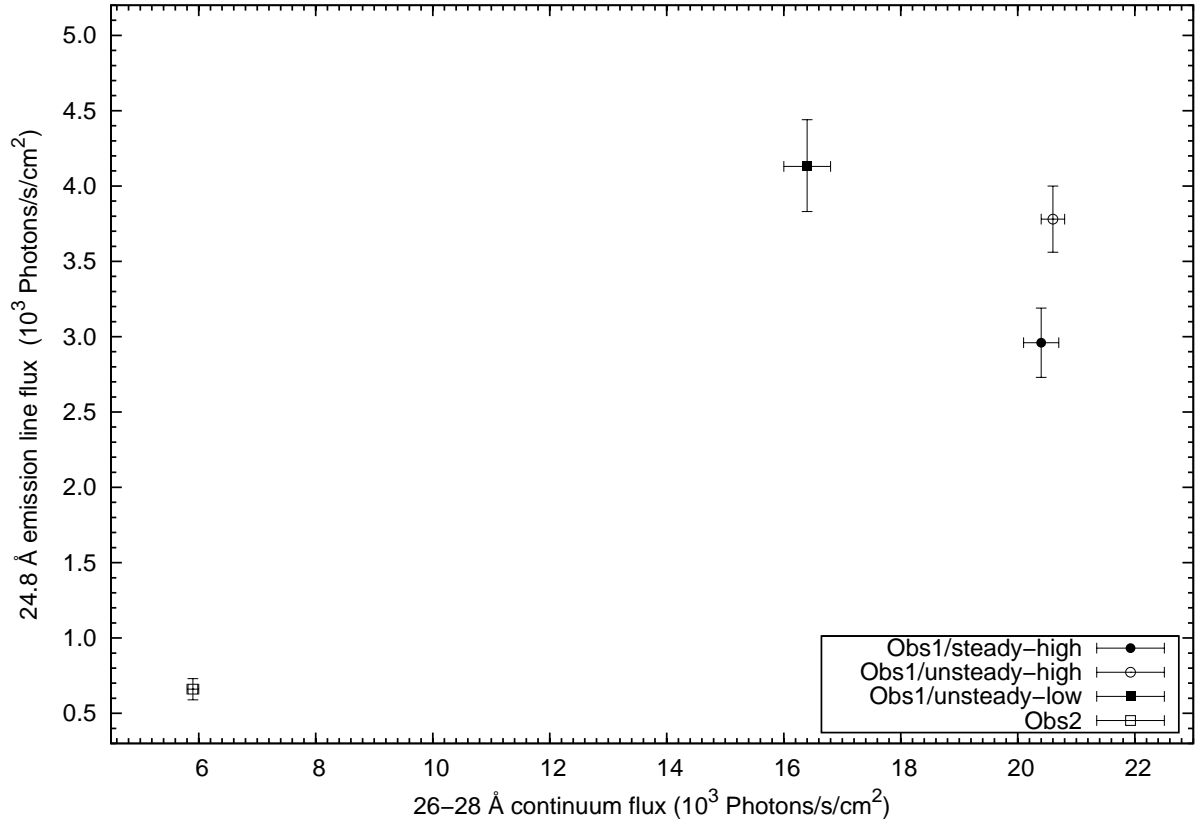}
\caption[]{%
Photon flux from the 24.8 \AA\ emission line vs. 26--28 \AA\ continuum flux
during the various observations/selections.
}
\label{f:eqw}
\end{figure}

While fitting, the normalization of RGS2 was allowed to vary w.r.t. that
of RGS1. The fits show that the RGS2 normalization factor is in the
range of 0.94--0.97.  Fig.~\ref{f:eqw} shows the 24.8 \AA\ line flux
and 26--28 \AA\ continuum flux for the various observations/selections.
Note that the ratio of line-to-continuum flux during Obs2 (0.11$\pm$0.03)
is similar to that during the steady-high state of Obs1 (0.14$\pm$0.02).
This strongly suggests that the line and the continuum fluxes are causally
linked and both originate from the accreting system (therefore ruling
out the possibility that the line is from some other astronomical source
in the angular proximity of \src; furthermore there is no other source
of comparable X-ray brightness in the EPIC image).

\section{Abundance estimates using XSTAR photoionization code} 
In order to obtain a self-consistent understanding of the elemental
abundances in the observed spectra we used the XSTAR photoionization
code \citep{xstarref}.  A grid of 480 XSTAR models was created from the
following parameters:
(1) model column density of the emitting plasma was sampled 
between $10^{19-24}$ cm$^{-2}$, 
(2) the ionization parameter (log($\xi$) where $\xi$=$L/nR^2$) was 
sampled between 0--4, 
(3) N abundance was sampled between 1--100 times relative to solar,
and
(4) O abundance sampled between 0.01--1 times relative to solar.  
A hydrogen nucleus density of 10$^{12}$ cm$^{-3}$, and a covering
fraction of 0.5 were assumed.  A high turbulent velocity of 1,000 km
s$^{-1}$ was used to simulate the line broadening.  Phenomenological
blackbody+power law fit to the 0.6--10 keV EPIC-PN data obtained during
the Obs1/steady-high state ($kT_{\rm BB}$=0.86 keV, $N_{\rm BB}$=147,
$\Gamma_{\rm PL}$=2.20, $N_{\rm PL}$=0.42) were used as user-defined
incident radiation field in XSTAR.  For a source distance of 1 kpc
(note that \citealt{Welsh+2010} reported a cloud of \ion{Na}{1} and
\ion{Ca}{2} at \about100 pc in the direction of \src; moreover, given the
high galactic longitude of \src, a distance of $\gg$1 kpc would put the
source far out of the galactic plane), the flux measured by PN implies
a luminosity of 2.6$\times$10$^{35}$ erg s$^{-1}$.  The remaining XSTAR
parameters were set to their default values.  An XSPEC table model
created from this grid was fit to Obs1/steady-high state.  The best
fit suggests a model column density of 3.2$\times$10$^{21}$ cm$^{-2}$
and log($\xi$)=2.3.  The 90\% confidence lower limit on log($\xi$) is 2.1
(mainly based on an upper limit on the \nvi/\nvii\ ratio), but the upper
limit on log($\xi$) could not be constrained from the data.  Since there
is only one statistically significant narrow line in the spectrum, it is
not possible to constrain the abundance of any single element.  Rather,
the fits are only able to give limits on the relative N/O ratio. The
fits suggest N/O overabundance $>$\NOsolarval\ ($>$\NOsolarll\ at 90\%
confidence) with respect to solar. This translates to an absolute N/O
of $>$\NOval\ ($>$\NOll\ at 90\% confidence) since we use the solar
abundance table of \citet{angr}.

We also created XSTAR models assuming the line is due to redshifted
\oviii\ Ly-$\alpha$ and solar nitrogen abundance.  The best-fit to this
redshifted \oviii\ model with $\chi^2$/$\nu$=4702/3608 is statistically
slightly less favored than the N-rich model with $\chi^2$/$\nu$=4673/3607.
We note that the best-fit log($\xi$) was pegged at the maximum value
allowed by our redshifted oxygen table model, implying a rather high
log($\xi$)$>$6 would be needed in that scenario.

\section{Discussion and Conclusions} \label{s:conclusion} 
\src\ is one of the few exceptional XRBs whose low-energy X-ray spectrum
can be studied in detail due to low interstellar extinction.  Fits to its
RGS spectra imply a column density of (2.1--4.6)$\times10^{20}$ cm$^{-2}$,
which is at least an order of magnitude below most XRBs.  Both \xmm\
observations of \src\ show a strong emission line near 24.8 \AA.  If the
line is from \nvii, this would require the donor to have an extremely high
N/O abundance ($>$\NOsolarval) relative to solar, based on the weakness
of oxygen lines of similar charge states.  Typically, for an XRB with a
solar-type donor, the oxygen abundance is \about8$\times$ greater than
that of nitrogen, and the X-ray spectra show lines from H- and He-like
ions of oxygen.  Even in the intermediate-mass X-ray binary Her X-1,
wherein N/O is 4$\times$ solar, O lines are observed to be as strong
or much stronger than N lines from similar charge states \citep[see,
e.g.,][]{Jimenez-Garate+2002}.  While N overabundance is not uncommon in
low-mass XRBs (e.g., in 4 of 5 UV spectra analyzed by \citet{Raymond1993}
the N/C ratios were several times solar, with a 9:1 ratio in Cyg X-2
being highest), the extremely high N/O in \src\ makes it an unique XRB.

Given the lack of source distance and any signature of orbital periodicity
so far, it is only possible to speculate on the nature of the donor
star in this system based on its strong N/O overabundance and color (see
below), and exotic stars such as hot, core-helium burning subdwarfs (sdB,
sdO), or degenerate white dwarfs (WD) appear to be strong candidates.
Hot subdwarfs were first discovered at high Galactic latitudes by
\citet{HumasonZwicky1947}, and subsequent studies have shown that about
half of them are in binary systems with WD or low-mass main sequence
stars. It is generally thought that the hydrogen-rich outer envelope
of the progenitor hot subdwarf in a binary system is lost via either
Roche-lobe overflow or common-envelope ejection mechanism, thus exposing
CNO products dredged up from the core \citep[see, e.g.,][]{Han+2002}.
\citet{Naslim+2010} have recently determined abundances for 6 sdB stars.
The absolute N/O ratios were 10, 13 and 20 for three of them (SB~21,
LB~1766, and BPS~CS~22940--0009), and oxygen was not detected in the
other three.  These stars were Ne-rich as well, so future observations of
the Ne spectrum of \src\ would be especially interesting.  One particular
sdB star, PG 1219+534, is observed to have N/O$\gtrsim$100 \citep[see,
e.g.,][for a review]{Heber2009}.  Based on recent evolutionary
calculations studying formation of ultracompact binaries (UCBs)
by \citet{Nelemans+2010}, it is also possible that the donor is a
helium WD. In the absence of detection of any C lines, N/C ratio is
unconstrained but likely high given the strength of the \nvii\ line.
Based on the diagnostics given by \citet{Nelemans+2010}, a high N/C
would also point towards a helium WD.

Assuming that the magnitudes reported in the USNO-B1.0 catalog for the
optical counterpart were made during quiescence, the temperature inferred
from the de-reddened B-R color is $\sim$46,000 K. This almost certainly
rules out typical donors found in low-mass XRBs, which tend to be of
late spectral type (K and M) and hence cooler.  The high temperature
suggested by the USNO photometry would also support a hot subdwarf donor
scenario. A caveat to the above scenario is that the mass transfer rate
during quiescence may also be non-negligible, so that there exists an
accretion disk even during X-ray quiescence (making the optical colors
appear blue).  Also,  He-sdOs are thought to evolve from merger of WDs,
and it may be difficult to place such a merger product in a binary with
a NS.  The low column density towards the source makes it unlikely that
the donor is a Wolf-Rayet star because Wolf-Rayet stars drive massive
winds creating high internal columns, e.g., as seen in Cyg X-3.

If the donor is indeed a WD/sdB/sdO, then the system is likely an UCB with
orbital period ranging from a few tens of minutes to $\sim$ few hours.
Observations of such UCBs are sparse, and their evolution is only recently
being explored thoroughly \citep[see, e.g.,][]{TaurisvdHeuvel2006,
Nelemans+2010}.  From a sample of 50 sdB binaries \citet{Geier+2008} found
that possibly four could have a neutron star or black hole donor. However,
despite a sensitive search, no radio pulsations have yet been detected
from these four systems \citep{Coenen+2011}.  If the donor in \src\
is a hot subdwarf, this would be the first direct evidence of a binary
system with a hot subdwarf donor and a compact (black hole or neutron
star) accretor.

Another intriguing possibility, assuming solar-abundance plasma, is that
the observed line is a gravitationally redshifted \oviii\ Ly-$\alpha$
line (rest-frame $\lambda$=18.967 \AA) originating from the surface of
the NS.  A photon emitted with a rest-frame energy of $E_0$ from the
surface of a slowly rotating, spherically symmetric NS of mass $M_{\rm
NS}$ and radius $R_{\rm NS}$ is observed by an observer at infinity
to have an energy $E$ where $E/E_0=\sqrt{1-(2G/c^2)(M_{\rm NS}/R_{\rm
NS})}$.  Based on the most precise mass estimates from pulsar timing
experiments, observed NS masses range between 1.25--2\msun\ \citep[see,
e.g.,][]{KramerWex2009, Demorest+2010}.  Assuming the observed 24.8
\AA\ line in \src\ is gravitationally redshifted \oviii\ Ly-$\alpha$,
the above mass range would imply a radius of 8.9--14.2 km.  In this
redshifted oxygen line scenario, the width of the line could be used
to put an upper limit on the relative size of the emission region as
follows: assume that the width of the observed line profile is mainly due
to superposition of lines originating at various heights above the NS.
Since a photon emitted from the surface (i.e. from a radius of R$_{\rm
NS}$) is more redshifted than a photon emitted from R$_{\rm NS}$+$\Delta
R$, the observed width of the narrow line may be related to $\Delta R$ as
$\Delta R/R_{\rm NS} \sim \sigma/(E_0-E) \sim 0.023$.  This is obviously
an upper limit to $\Delta R$, since it ignores any bulk motion.  However,
a redshifted \oviii\ line scenario would also require other redshifted
lines of oxygen to be present in the spectra, which are not seen.  Also,
the hint of \nvi\ resonance and intercombination lines favors the N/O
overabundance scenario.  Planned high-resolution optical/UV spectroscopy
will be key in understanding the nature of this unique source.

\acknowledgments
We thank the anonymous referee for helpful comments.  It is a pleasure
to thank Norbert Schartel and the \xmm\ planning team for carrying out
the ToO observations.  We would also like to acknowledge the use of the
daily monitoring data obtained by the RXTE which has provided excellent
all-sky coverage of X-ray sources over the past 15 years.


\begin{deluxetable}{lccccc}
\tabletypesize{\scriptsize} 
\tablewidth{0pt}
\tablecaption{Best fit continuum parameters for the RGS data\tablenotemark{1}
	       \label{t:cont_fits}}
\tablehead{
  \colhead{Obs/Selection} &
  \colhead{\nh$\times10^{20}$} &
  \colhead{$\Gamma_1$} & \colhead{$E_{\rm b}$} &
  \colhead{$\Gamma_2$} & \colhead{$N$}\\
   & (cm$^{-2}$) & & (keV) & &
} 
\startdata
Obs1/steady-high & 
 $3.88_{-0.74}^{+0.71}$ &	
 $1.87_{-0.28}^{+0.28}$ &	
 $0.475_{-0.012}^{+0.019}$ &	
 $1.20_{-0.05}^{+0.05}$ &	
 $0.203_{-0.031}^{+0.035}$ \\ \\	
Obs1/unsteady-high & 
 $2.29_{-0.25}^{+0.25}$ &	
 $1.18_{-0.03}^{+0.03}$ &	
 $1.491_{-0.048}^{+0.050}$ &	
 $1.60_{-0.15}^{+0.18}$ &	
 $0.303_{-0.003}^{+0.003}$ \\ \\	
Obs1/unsteady-low & 
 $3.57_{-0.65}^{+0.18}$ &	
 $2.58_{-0.98}^{+0.42}$ &	
 $0.396_{-0.014}^{+0.022}$ &	
 $1.57_{-0.05}^{+0.05}$ &	
 $0.079_{-0.028}^{+0.113}$ \\ \\	
Obs2 & 
 $3.33_{-0.46}^{+0.73}$ &	
 $2.09_{-0.56}^{+0.91}$ &	
 $0.396_{-0.030}^{+0.050}$ &	
 $1.31_{-0.04}^{+0.05}$ &	
 $0.041_{-0.026}^{+0.028}$ \\	
\enddata
\tablenotetext{1}{
Errors are 90\% confidence limits.
Col. (1) gives the observation and time/rate selection as discussed in \S\ref{s:data}; 
col. (2) is the best-fit \nh\ in units of 10$^{20}$ atoms cm$^2$; 
cols. (3,5) are power law photon indices for $E<E_b$ and $E>E_b$ respectively; 
col. (4) is the break point for the energy; 
col. (6) is the power law normalization at 1 keV, i.e., photons keV$^{-1}$ cm$^{-2}$ s$^{-1}$ at 1 keV.;
}
\end{deluxetable}

\begin{deluxetable}{lccccc}
\tabletypesize{\scriptsize} 
\tablewidth{0pt}
\tablecaption{Best fit narrow line parameters for the RGS 
 data\tablenotemark{1}
	       \label{t:line_fits}}
\tablehead{
  \colhead{Obs/Selection} &
  \colhead{Center} & \colhead{$\sigma$} & \colhead{10$^3\times$Flux} & \colhead{Equiv. width} & $\chi^2/\nu$\\
   & (\AA) & (\AA) & (photons cm$^{-2}$ s$^{-1}$) & (m\AA)
} 
\startdata
Obs1/steady-high & 
 $24.801_{-0.017}^{+0.018}$ &	
 $0.168_{-0.018}^{+0.020}$ &	
 $2.96_{-0.23}^{+0.23}$ &	
 $190_{-15}^{+15}$ & 		
 $4650.2/3609$ \\ \\ 
Obs1/unsteady-high & 
 $24.800_{-0.014}^{+0.014}$ &	
 $0.211_{-0.022}^{+0.026}$ &	
 $3.78_{-0.21}^{+0.22}$ &	
 $273_{-15}^{+16}$ & 		
 $5138.0/3902$ \\ \\ 
Obs1/unsteady-low & 
 $24.824_{-0.016}^{+0.016}$ &	
 $0.175_{-0.023}^{+0.026}$ &	
 $4.13_{-0.30}^{+0.31}$ &	
 $343_{-25}^{+26}$ & 		
 $2632.1/2575$ \\ \\ 
Obs2 & 
 $24.738_{-0.029}^{+0.029}$ &	
 $0.219_{-0.038}^{+0.053}$ &	
 $0.66_{-0.07}^{+0.08}$ &	
 $155_{-17}^{+18}$ & 		
 $4213.0/3580$ \\ 
\enddata
\tablenotetext{1}{
Errors are 68\% confidence limits.
Col. (1) same as in Table~\ref{t:cont_fits}; 
cols. (2--5) best fit Gaussian line center, width ($\sigma$), line flux, and
equivalent width for the narrow line; 
col. (6) $\chi^2$/d.o.f for the best fit to the 7--36 \AA\ RGS data.
}
\end{deluxetable}

\begin{figure} 
\centering
\includegraphics[angle=-90, width=1.0\textwidth]{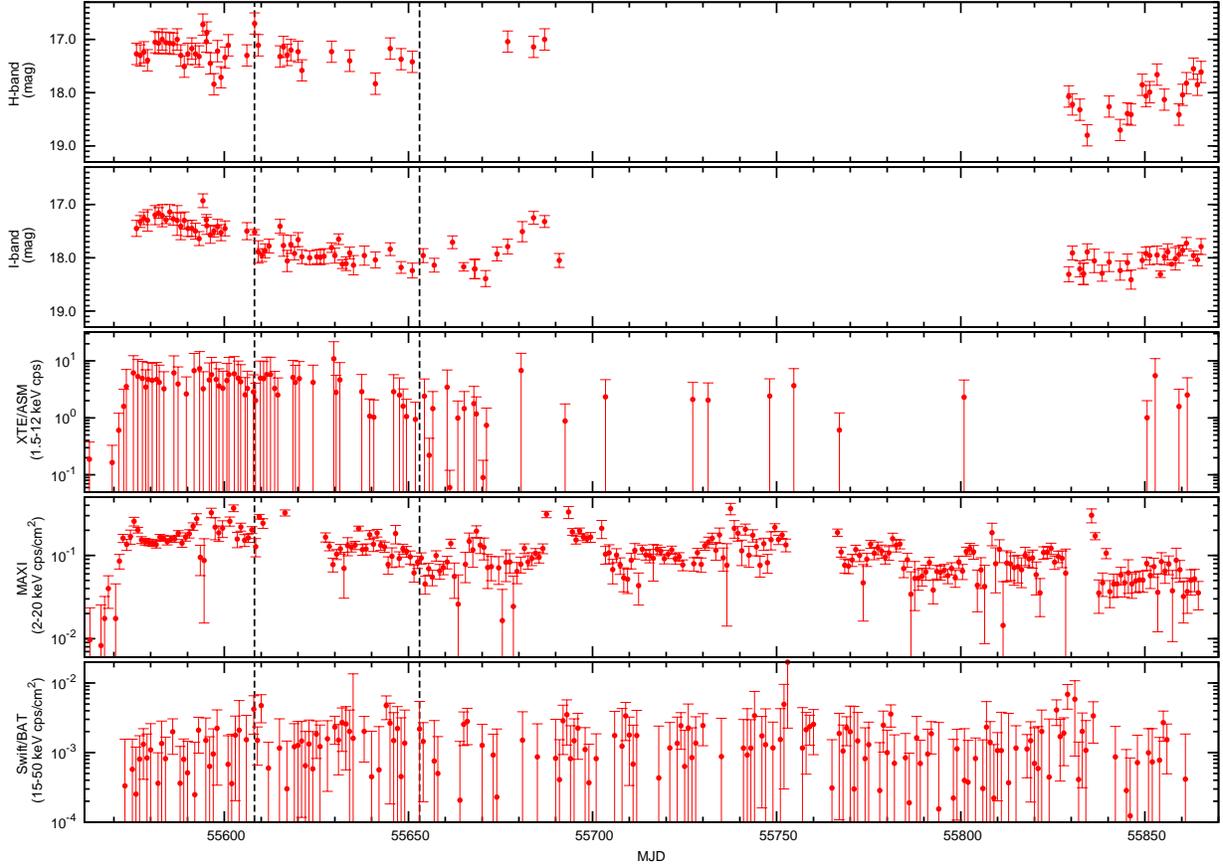}
\caption[]{%
Supplementary figure showing the multi-wavelength monitoring of
\src. The top two panels show H- and I-band light curve obtained with
the CTIO/SMARTS 1.3m telescope. The next three panels show the 1.5--12
keV RXTE/ASM light curve (third panel from top), the 2--20 keV MAXI light
curve (fourth panel from top), and the 15--50 keV Swift/BAT light curve
(bottom panel).
}
\label{f:monitoring}
\end{figure} 

\end{document}